\documentclass[twocolumn,aps,prd,graphics,floatfix,a4paper,superscriptaddress,showkeys,showpacs]{revtex4-1} % {revtex4}

\usepackage{CJK}
\usepackage{graphicx}
\usepackage{mathrsfs}
\usepackage{bm}
\usepackage{amsmath}
\usepackage{dcolumn}
\usepackage{epstopdf}
\usepackage{dsfont}
\usepackage{amssymb}
\usepackage{tabularx}
\usepackage{array}
\usepackage{float}
\usepackage{color}
\usepackage{epstopdf}
\usepackage{mathrsfs}
\usepackage[colorlinks, linkcolor=blue,anchorcolor=blue,citecolor=blue,urlcolor=blue]{hyperref}
\usepackage{appendix}
\usepackage{lipsum}
\usepackage{makecell}
\usepackage{supertabular}
\usepackage{threeparttable}

\begin{document}

\title{Two-dimensional topological semimetals}% Force line breaks with \\
%	\thanks{A footnote to the article title}%
	
\author{Xiaolong Feng}
\address{Research Laboratory for Quantum Materials, Singapore University of Technology and Design, Singapore 487372, Singapore}%Lines break

\author{Jiaojiao Zhu}
\address{Research Laboratory for Quantum Materials, Singapore University of Technology and Design, Singapore 487372, Singapore}

\author{Weikang Wu}
\email{weikang.wu@ntu.edu.sg}
\address{Research Laboratory for Quantum Materials, Singapore University of Technology and Design, Singapore 487372, Singapore}
\address{Division of Physics and Applied Physics, School of Physical and Mathematical Sciences, Nanyang Technological University, Singapore 637371, Singapore}

\author{Shengyuan A. Yang}
\address{Research Laboratory for Quantum Materials, Singapore University of Technology and Design, Singapore 487372, Singapore}

%\date{*}% It is always \today, today,
	
\begin{abstract}
The field of two-dimensional topological semimetals, which emerged at the intersection of two-dimensional materials and topological materials, have been rapidly developing in recent years. In this article, we briefly review the progress in this field. Our focus is on the basic concepts and notions, in order to convey a coherent overview of the field. Some material examples are discussed to illustrate the concepts. We discuss the outstanding problems in the field that need to be addressed in future research.
\end{abstract}
	

\maketitle

\section{\label{intro}Introduction}

Two-dimensional (2D) materials and topological materials have been two most active research fields in the past decade.
The field of 2D materials was kicked off with the successful realization of graphene in 2004~\cite{Novoselov2004}. Since then, many new atomic-thick 2D materials have been achieved in experiment, and much more have been predicted in theory. For example, a 2D material database in 2019 documented more than 6,000 monolayer materials that can in principle be realized~\cite{Zhou2019}.

On the other hand, although the history of topological states of matter can be traced back at least to the study of quantum Hall effect in the 1980s~\cite{Klitzing1980,Thouless1982,Prange1990}, interestingly, the field truly flourished also around 2004, accompanying the birth of graphene. In two consecutive works~\cite{Kane2005,Kane2005Quantum}, Kane and Mele proposed the concept of 2D topological insulators (also known as quantum spin Hall insulators), and suggested graphene as a candidate material. The topological concepts were firstly discussed in insulators~\cite{Hasan2010,Qi2011,Shen2012,Bernevig2013}. Later, in 2011, the proposal of  Weyl semimetal concept by Wan \emph{et al.}~\cite{Wan2011Topological} motivated extensive research on topological semimetals (TSMs)~\cite{Chiu2016Classification,Bansil2016,Armitage2018Weyl}. For this sub-field, the research was initially on three-dimensional (3D) systems, but soon, it was extended to 2D and closely interacted with the field of 2D materials.

In fact, in retrospect, graphene itself represents a good example of a 2D TSM --- the 2D Weyl semimetal, in the absence of spin-orbit coupling (SOC). The key features of a TSM can be illustrated using graphene. In the band structure of a TSM, near Fermi level, there exist protected band degeneracy points. In graphene, these are the linear nodal points at $K$ and $K'$ points of the Brillouin zone (BZ). Because of these points, the low-energy electron quasiparticles exhibit distinct features, such as unusual dispersions, pseudospin structures, and etc~\cite{Neto2009}. In graphene, the low-energy electrons mimic 2D Weyl fermions. (In literature, they were also often called the Dirac fermions. We will clarify these concepts later.) Consequently, TSMs can manifest exotic physical properties. Many remarkable properties of graphene, such as ultrahigh mobility, Klein tunneling effect, and weak antilocalization, are tied with its topological character~\cite{Neto2009}. Actually, the clean band structure and large linear dispersion window make graphene the best example of all available TSM materials to date.

\begin{figure}[t]
	\centering
	\includegraphics[angle=0, width=0.47\textwidth]{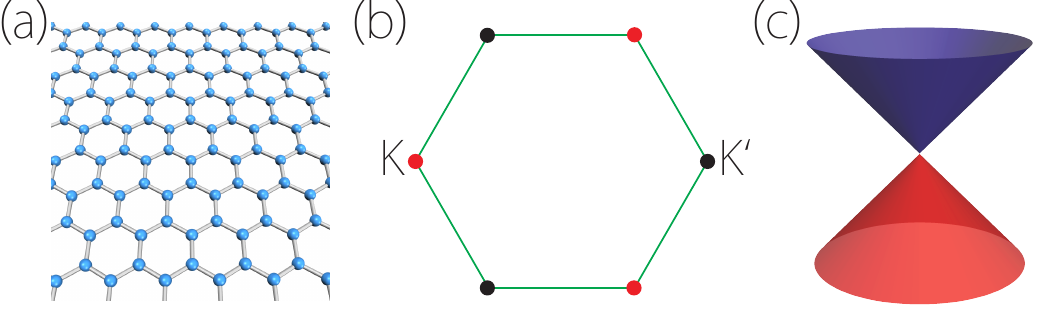}
	\caption{(a) Lattice structure of graphene, consisting of a single sheet of carbon atoms arranged in a honeycomb lattice. (b) At Fermi level, graphene has two linear Weyl points located at the corners $K$ and $K'$ of the BZ. (c) shows the Weyl-cone dispersion around a Weyl point.}
	\label{graphene}
\end{figure}

Hence, we see that the two fields, TSMs and 2D materials, are intertwined from the very beginning. Over the past ten years, many new 2D TSM phases have been proposed and some of them have found realization in real 2D materials. Sitting at the intersection of two fields, the rapid progress of 2D TSMs gains impetus from both fields. Particularly, the high controllability of 2D materials is a great advantage for fundamental research as well as practical applications.

In this article, we will review the recent progress on 2D TSMs.  Due to the enormous body of literature on this subject, we definitely cannot cover all the important topics in this short review. We will try to focus on the basic concepts and notions to provide an overview of the field. And we apologize in advance for possible omission of some relevant works.

This review is organized as following. In Sec.~\ref{basicConcept}, we introduce the different schemes for classifying TSM states and present the standards for a good TSM material. In Sec.~\ref{tnpsm} and Sec.~\ref{tnlsm}, we discuss 2D nodal-point and nodal-line TSMs, respectively, in nonmagnetic systems. We extend the discussion to magnetic systems in Sec.~\ref{magtsm}. In Sec.~\ref{tes}, we review the topological edge states in 2D TSMs. An outlook of the field is presented in Sec.~\ref{outlook} .

\section{\label{basicConcept} Classify 2D TSM states}

The key feature of a TSM is the band degeneracy near the Fermi level, so the schemes to classify TSMs are based on properties of such degeneracy. Below, we lay out these properties.

\begin{figure}[t]
	\centering
	\includegraphics[angle=0, width=0.47\textwidth]{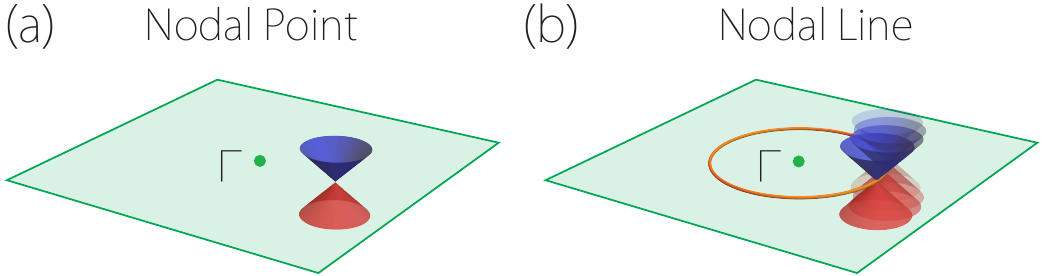}
	\caption{Classification based on dimensionality. (a) 0D nodal points. (b) 1D nodal line.}
	\label{dim}
\end{figure}

{\color{blue}\emph{Dimensionality.}} This refers to the dimensionality of the degeneracy. For 2D systems, there are two possibilities: the band degeneracies may either form 0D nodal points or 1D nodal lines. Correspondingly, we have nodal-point TSMs and nodal-line TSMs. This is the primary classification scheme.

{\color{blue}\emph{Number of degeneracy.}} This tells how many bands are degenerate at the same momentum and same energy. It is at least twofold, which is also the mostly encountered case. In the field of TSMs, twofold degeneracy is conventionally referred to as ``Weyl", while fourfold degeneracy is referred to as ``Dirac". This naming stems from the analogy with Weyl and Dirac fermions~\cite{Dirac1928,Weyl1929} in relativistic quantum field theory.

\begin{figure}[h]
	\centering
	\includegraphics[angle=0, width=0.45\textwidth]{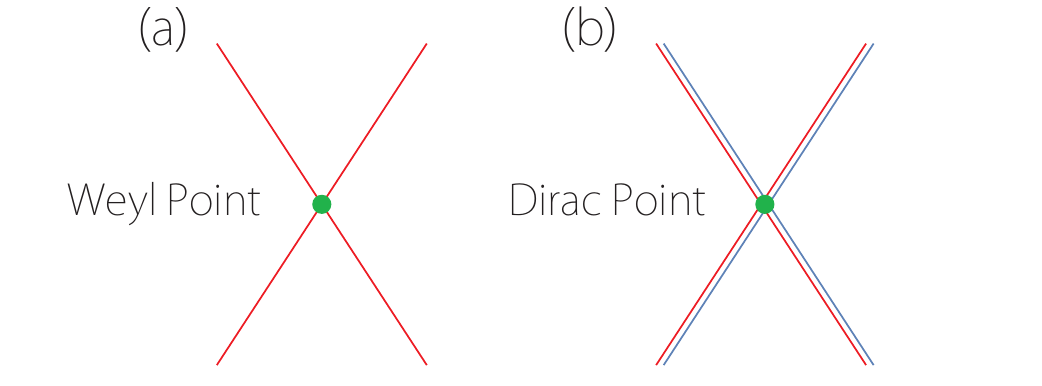}
	\caption{Classification based on the number of degeneracy. (a) Weyl point, which is twofold degenerate. (b) Dirac point, which is fourfold degenerate.}
	\label{deg}
\end{figure}

It should be noted that in many works on graphene, the nodal points are called ``Dirac". However, the correct naming should be ``Weyl", because the system is essentially spinless and then each point is twofold degenerate and described by the Weyl model in 2D. The misuse of ``Dirac" here is a historical issue: In early works on graphene, it was common to combine the two Weyl models at the two nodal points $K$ and $K'$ into a $4\times 4$ Dirac-like form, hence it was called Dirac~\cite{Yang2016}. This usage is clearly inconsistent with the later conceptual development in the TSM field, where the naming should reflect the degeneracy at the \emph{same} $k$ point.

{\color{blue}\emph{Type of dispersion.}} In the study of 3D Weyl semimetals, a Weyl point is classified as type-I if the Weyl cone (formed by the two crossing bands) is up-right along any momentum direction around the point; and as type-II if this is not the case, namely, in at least one direction, the cone is tipped over~\cite{Soluyanov2015,Xu2015}. The reason for this classification is that the two types correspond to distinct Fermi surface topologies: The Fermi surface around a type-I point consists of a single pocket and it may shrink into a Fermi point if the Weyl point is exactly at the Fermi energy; in contrast, around a type-II point, there are co-existing electron and hole pockets, and the two pockets touch at the Weyl point if the point is at the Fermi energy.

\begin{figure}[h]
	\centering
	\includegraphics[angle=0, width=0.45\textwidth]{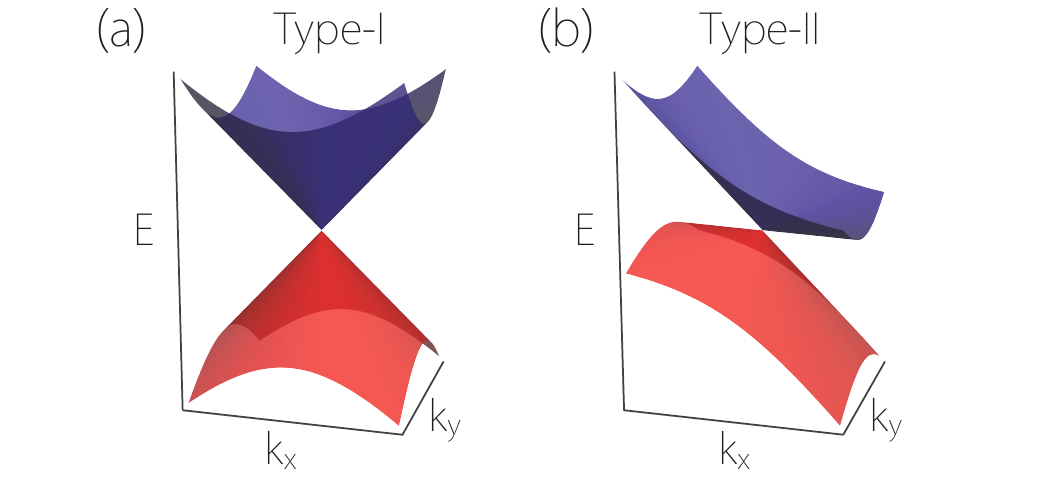}
	\caption{Two types of dispersion for a nodal point. (a) Type-I dispersion. (b) Type-II dispersion.}
	\label{type}
\end{figure}

The classification can be extended to nodal lines, as first proposed by Li \emph{et al.} in Ref.~\cite{Li2017}. Besides type-I and type-II nodal lines, there is a third possibility: the hybrid nodal lines~\cite{Li2017}. The different classes also have distinct Fermi surface topologies, leading to unique physical properties~\cite{Li2017,Zhang2018}.

A recent review of this classification can be found in Ref.~\cite{Li2020a}. Hence, we shall not elaborate further on this point.

{\color{blue}\emph{Order of dispersion.}} In most cases, degeneracies are formed by linear crossing between bands.  Nevertheless, with the participation of certain crystal symmetries, the linear dispersion term around a degeneracy may be killed, which makes the higher-order term to become the leading order. This order is an important character, as it determines the scaling of density of states, pseudospin structure, and topological charge, and etc.

\begin{figure}[h]
	\centering
	\includegraphics[angle=0, width=0.45\textwidth]{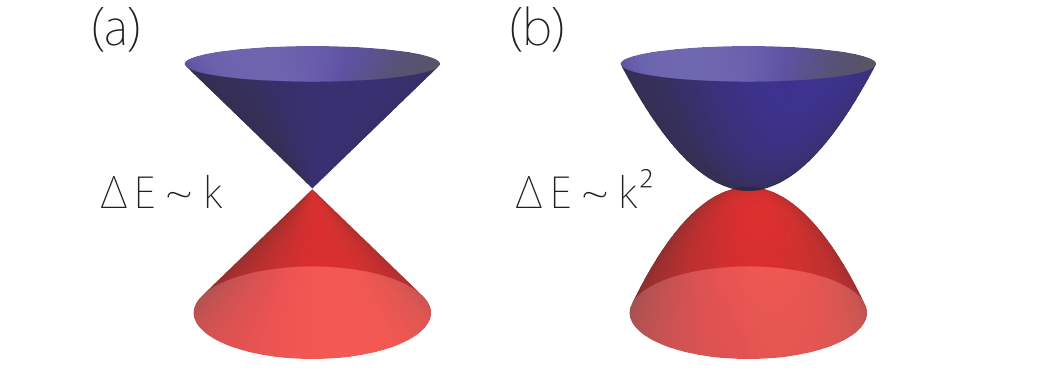}
	\caption{Order of dispersion for a nodal point. (a) Conventional linear dispersion. (b) Quadratic dispersion. $\Delta E$ is the energy splitting between the two bands.}
	\label{disp}
\end{figure}

In 3D systems, higher-order nodal points have been extensively studied~\cite{Xu2011,Fang2012,Yang2014a,Gao2016,Wu2020a,Yu2021}. In recent works, higher-order nodal lines have also been reported~\cite{Yu2019,Zhang2021}. In 2D, higher-order lines appear unlikely, but higher-order points do exist. For example, twofold nodal points with quadratic dispersion, called double Weyl points, have been reported in several 2D materials~\cite{Zhu2016Blue,Wang2018,Hua2020tunable}.

{\color{blue}\emph{Other features.}} Apart from the main schemes above, there are other features that classification of 2D TSMs can be based on. For example, nodal points can be classified based on their locations. In Ref.~\cite{Lu2016Multiple} by Lu \emph{et al.}, 2D Weyl points located on high-symmetry path or generic $k$ points are referred to as ``unpinned", as their counterparts at high-symmetry points (like those in graphene) are pinned by symmetry and cannot freely move.

\begin{figure}[h]
	\centering
	\includegraphics[angle=0, width=0.45\textwidth]{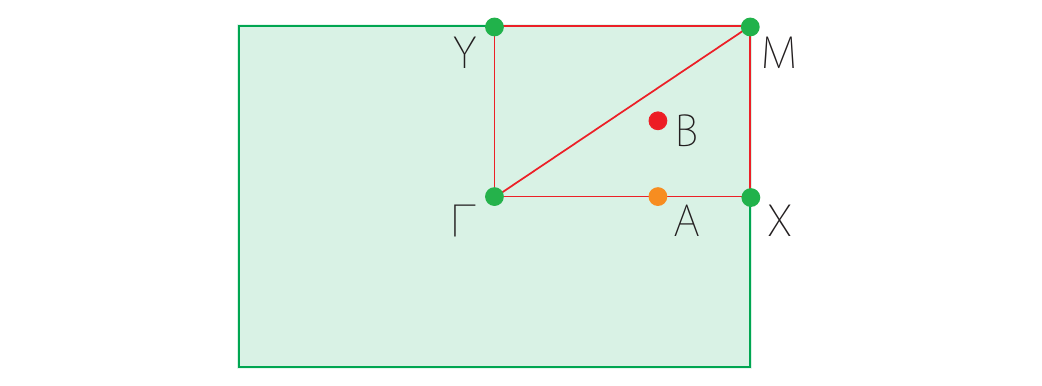}
	\caption{Nodal points at high-symmetry points (green dots) are pinned. Nodal points on high-symmetry paths (yellow dot) and at generic $k$ point (red dot) are unpinned. }
	\label{loc}
\end{figure}

This classification based on the location in the BZ is connected (but not identical) to the concepts of accidental and essential band degeneracies. ``Essential" means the presence of the degeneracy is guaranteed by symmetry, and it cannot be removed unless the symmetry is broken.
In contrast, the presence of accidental degeneracies further requires band inversion, and they can be removed while preserving the symmetry. Nodal points on high-symmetry paths or generic $k$ points, such as the unpinned points in Ref.~\cite{Lu2016Multiple}, are typically accidental.

\begin{figure}[h]
	\centering
	\includegraphics[angle=0, width=0.45\textwidth]{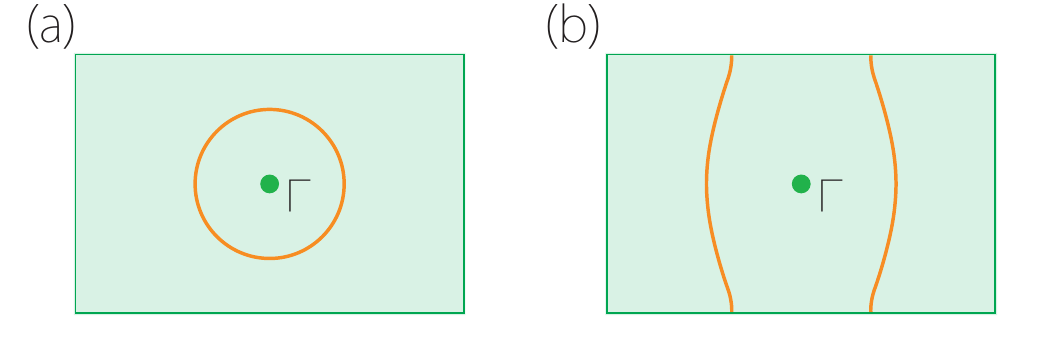}
	\caption{Classification of nodal loops based on the topology in the BZ. (a) An isolated nodal ring can continuously shrink into a point. (b) Each individual nodal loop here cannot shrink into a point by itself, as it traverses the BZ.  }
	\label{patt}
\end{figure}

Nodal lines may also be classified according to its pattern in the BZ. At single-particle level, the continuity of band structure guarantees that nodal lines must form closed loops. Hence, the terminologies of ``nodal line" and ``nodal loop" are used interchangeably. Since the BZ has the topology of a torus which is not simply connected, a loop in the BZ can be classified according to whether or not it can shrink into a point. This classification scheme was first proposed in Ref.~\cite{Li2017}. For 2D systems, a nodal loop can be characterized by two integers, each representing its number of winding along a direction in the BZ. This characterizes the difference between the two loops shown in Fig.~\ref{patt}. The loop in Fig.~\ref{patt}(a), often called a nodal ring and centered around a high-symmetry point, can continuously shrink into a point (while preserving the symmetry), but each individual loop in Fig.~\ref{patt}(b) cannot, because it traverses the BZ once.

Another important feature is regarding whether SOC is considered in the system or not, because it fundamentally affects the symmetry group representation (single \emph{vs.} double representations).
When SOC is absent, as in materials with light elements, bosonic and classical systems, the nodal points and lines may be called ``spinless". In the presence of SOC, the required symmetry condition for a degeneracy is generally more stringent, so the adjective ``spin-orbit" is often added to the naming.

Related to spin, works in the field also stress whether a TSM occurs in a magnetic system or not. This is because the presence/absence of time reversal symmetry has fundamental importance on the topological classification, and spin polarization in magnetic materials also impacts the physical properties. For magnetic TSMs, the most interesting case is when the low-energy window that contains the nodal points/lines belong to a single spin channel. Such states were known as the topological half semimetal, i.e., simultaneously a half metal (for spin) and a semimetal (for charge).

In the following sections, we shall further expose the concepts mentioned above with some realistic material examples.
Regarding materials, we note that a good TSM material should satisfy the following requirements.
\begin{itemize}
  \item The target band degeneracy should be close to the Fermi level, such that the nontrivial emergent fermions can play an important role in physical properties of the material.
  \item The low-energy band structure should be clean. The desired case is when the low-energy window contains only the target band degeneracy.
  \item The characteristic dispersion should dominate in a large energy window. For example, for a (linear) Weyl point, it is desired that the linear dispersion dominates over a large window around the Fermi level.
\end{itemize}

\section{\label{tnpsm} 2D nodal-point TSM}

In 2D, stable nodal points can only have twofold or fourfold degeneracy, corresponding to Weyl and Dirac types. Below, we shall discuss the different types separately.

{\color{blue}\emph{Spinless Weyl point.}} As mentioned, graphene is a 2D Weyl TSM. Its conduction and valence bands cross linearly at two Weyl points at $K$ and $K'$ points. Around the Weyl points, the electrons are described by the 2D Weyl model:
\begin{equation}\label{weyl}
  H_\tau=v_F(\tau k_x \sigma_x+k_y\sigma_y),
\end{equation}
where $v_F$ is the Fermi velocity, $\tau=\pm 1$ labels $K$/$K'$ (known as the valley index), and $\sigma$'s are the Pauli matrices. The energy and momentum here are measured from a Weyl point. Note that $\sigma$ is a pseudospin corresponding to the orbital degree of freedom (in the simple picture, these are the two $p_z$ orbitals at the two carbon atoms in a unit cell). Since the SOC is negligible in carbon materials, the real spin is a dummy degree of freedom here.

The two Weyl points have opposite ``chirality" determined by the sign of $\tau$. Here, chirality means the sense of rotation of the pseudospin of the valence band along a small circle surrounding the point: one is clockwise whereas the other is counterclockwise. When a gap $\Delta\sigma_z$ is opened by breaking the inversion symmetry, this chirality leads to valley-contrasting properties, like Berry curvatures and orbital magnetic moments, which form the basis for the existing paradigm of valleytronics~\cite{Xiao2007,Yao2008,Cai2013}.

The Weyl points in graphene are protected by multiple symmetries. For instance, the point corresponds to a 2D irreducible representation of the $D_{3h}$ little group at $K$/$K'$, and in fact the $C_3$ subgroup is sufficient for the protection. Besides, the protection also comes from the spacetime inversion symmetry $PT$. For a spinless system with $PT$, the Berry phase along a closed path $\ell$ must be $\pi$ quantized, which gives a $\mathbb{Z}_2$ valued 1D topological charge (also known as the first Stiefel-Whitney number)
\begin{equation}\label{1D}
  \nu=\frac{1}{\pi}\oint_\ell \bm{\mathcal{A}}(\bm k)\cdot d\bm k\quad \mod 2,
\end{equation}
where $\bm{\mathcal{A}}$ is the Berry connection for the valence bands. Each Weyl point in graphene carries a nontrivial $\nu=1$, hence it is stable against $PT$-invariant perturbations. In addition, the sublattice (or chiral) symmetry also gives a protection. The sublattice symmetry $S$ satisfies $\{S,H\}=0$. It is preserved on a bipartite lattice when the hopping occurs only between the two sublattices. As a manifestation of the sublattice symmetry, the energy spectrum is symmetric about $E=0$. Of course, in real materials, the spectra are never exactly symmetric.
Nevertheless, for many cases, $S$ still appears as a good symmetry in the low-energy window, and this is the case for graphene. Under $S$, the path $\ell$ has a $\mathbb{Z}$ valued topological charge $\nu_C$, which is just the $\nu$ in Eq.~(\ref{1D}) without taking the modulus 2. Hence, one Weyl point in graphene has charge $\nu_C=+1$, whereas the other has charge $-1$.

The Weyl electrons described in Eq.~(\ref{weyl}) endow graphene with many remarkable properties. Especially, the topological charge $\nu$ ($\pi$ Berry phase) directly leads to the weak anti-localization effect and unconventional Landau level structure~\cite{Neto2009}. More physical consequences can be found in the review articles on graphene~\cite{Neto2009}.

\begin{figure}[h]
	\centering
	\includegraphics[angle=0, width=0.45\textwidth]{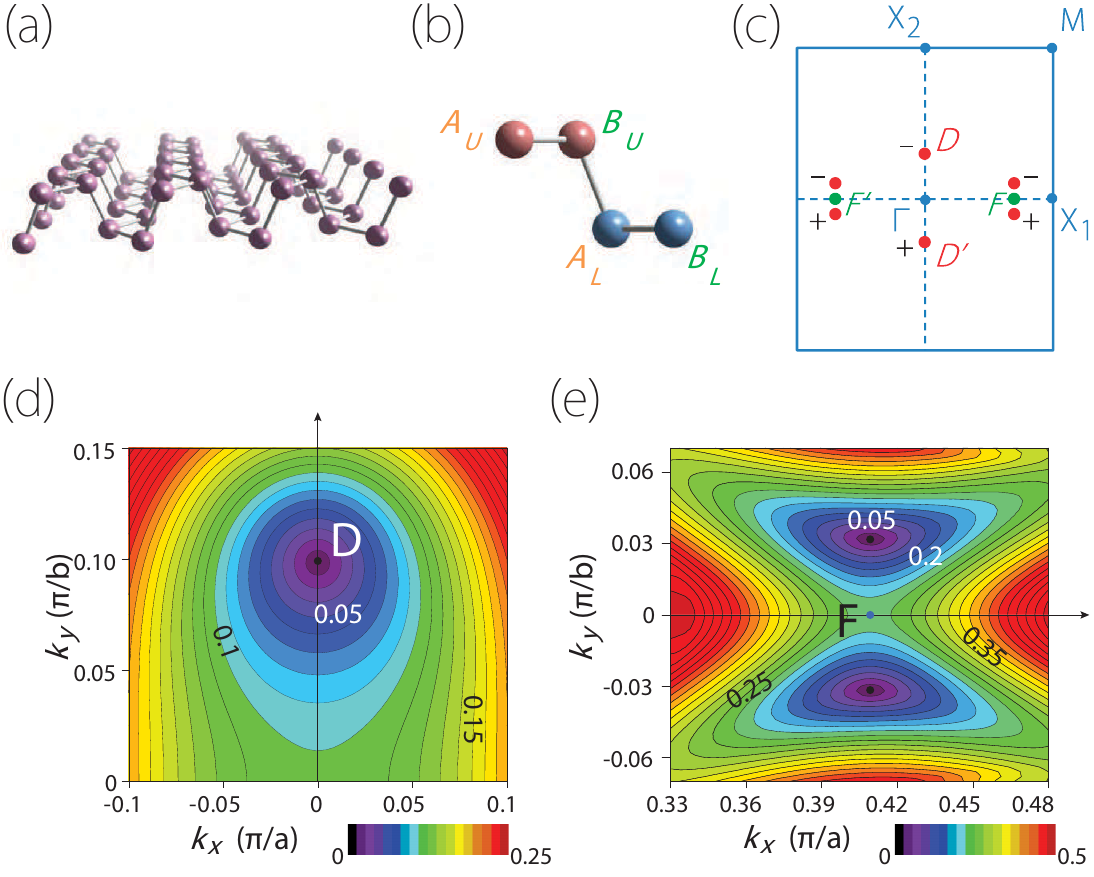}
	\caption{Unpinned Weyl points in group-Va monolayers with the phosphorene structure. (a) The lattice structure. (b) shows the unit cell with four-atom basis. (c) There are six Weyl points (red dots) in the BZ. Two are on high-symmetry path $\Gamma$-$X_2$. The other four are at generic $k$ points. (d) Anisotropic dispersion for a Weyl point $D$ on high-symmetry path. (e) Anisotropic dispersion for a pair of fully unpinned Weyl points around $F$ in (c). Figures adapted with permission from Ref.~\cite{Lu2016Multiple}.}
	\label{unpin}
\end{figure}

Similar spinless Weyl points have been identified in several other 2D materials, such as graphynes~\cite{Malko2012Competition}, some 2D boron allotropes~\cite{Zhou2014Semimetallic}, germanene~\cite{Liu2015Multiple}, and group-Va monolayers with the phosphorene structure~\cite{Lu2016Multiple}. It was noted that in some of these examples, the Weyl points appear on the high-symmetry path, like in $\beta$-graphyne~\cite{Malko2012Competition}, germanene on Al(111)~\cite{Liu2015Multiple}, and strained monolayer Na$_2$O~\cite{Hua2020tunable}. Moreover, in strained phosphorene structures, Weyl points at generic $k$ were reported~\cite{Lu2016Multiple}. These are the unpinned nodal points we mentioned in Sec.~\ref{basicConcept}. Compared to the pinned ones, they have two important differences. First, with symmetry-preserving perturbations, the unpinned points can move in the BZ. For points on the high-symmetry path, they can move along the path; for points at generic $k$, they can freely move in the BZ. As a result, the unpinned points may annihilate in pairs when they collide at the same point. Second, due to the reduced symmetry at the point, the dispersion around the point is anisotropic. For example, an effective model for an unpinned Weyl point on a high-symmetry path along $x$ may take the form of
\begin{equation}\label{tilt}
  H=w k_x+v_x k_x\sigma_x+v_y k_y\sigma_y.
\end{equation}
One notes that generally $v_x\neq v_y$, and there is a tilt term $wk_x$ which tilts the Weyl cone along $x$. For fully unpinned Weyl points at generic $k$, the dispersion is even more anisotropic. A model for such an example can be found in Ref.~\cite{Lu2016Multiple}.

\begin{figure}[h]
	\centering
	\includegraphics[angle=0, width=0.45\textwidth]{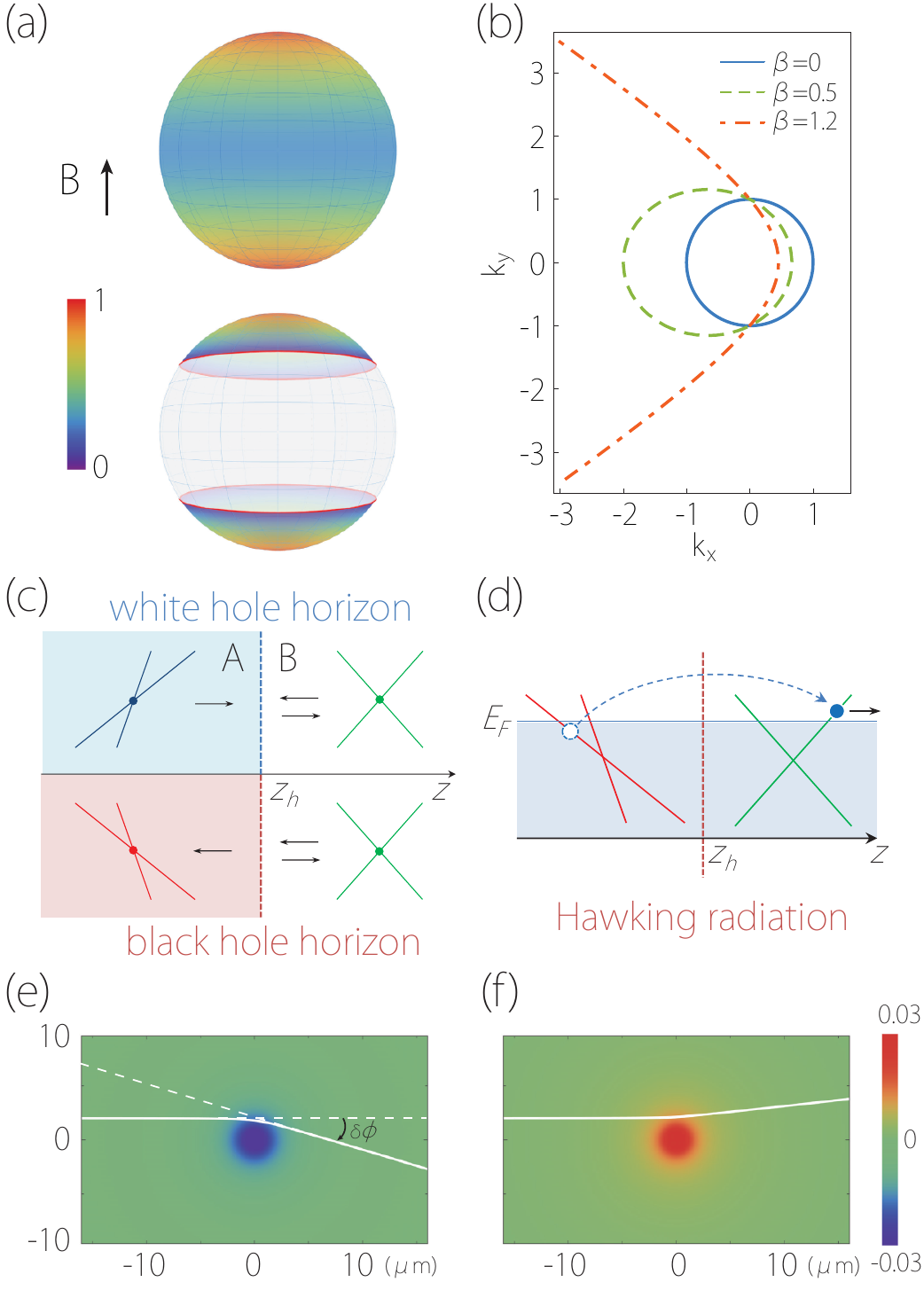}
	\caption{Some interesting effects for type-II dispersion. (a) Landau level squeezing factor plotted versus tilt direction on a unit sphere. Upper: Type-I point. Lower: Type-II point, in which the two red loops mark the critical angle where the Landau levels collapse. (b) Semiclassical orbit has a change in topology when going from type-I to type-II. (c) Schematic figure showing a white/black hole horizon simulated at an interface between type-I and type-II regions. (d) Analogue of Hawking radiation proposed in TSMs. (e-f) Analogue of gravitational lensing effect. The white line indicates a quasiparticle trajectory (geodesic) in a TSM under a non-uniform strain profile. Figures adapted with permission from Ref.~\cite{Yu2016,Guan2017}. }
	\label{unmag}
\end{figure}

It should be noted that for unpinned Weyl points, the cone is generally tilted, as in Eq.~(\ref{tilt}). When the tilt dominates the dispersion, for Eq.~(\ref{tilt}) this refers to $|w|>|v_x|$, the dispersion becomes type-II. The type-II dispersion can manifest in effects such as magnetic breakdown~\cite{OBrien2016}, Landau level collapse~\cite{Yu2016}, and magneto-optical response~\cite{Yu2016,Tchoumakov2016,Udagawa2016}. Due to the emergent Lorentz symmetry, by realizing a spatially varying tilt term, e.g., by lattice strain, one can simulate interesting effects from general relativity, such as event horizons, gravitational lens, and Hawking radiation, as proposed in Ref.~\cite{Guan2017}.

{\color{blue}\emph{Spin-orbit Weyl point.}} When SOC must be considered, the condition for a 2D Weyl TSM is more stringent. Since we need twofold Weyl points near the Fermi level, the bands must have spin splitting and the splitting should be large enough. Otherwise, bands with different spins would coexist in the low-energy window, and we cannot have a clean TSM band structure. In nonmagnetic materials, the splitting comes purely from SOC. In magnetic materials, exchange field typically plays the leading role in the splitting.

For nonmagnetic 2D materials, a general consideration is that the inversion symmetry $P$ must be broken. Otherwise, each band is twofold degenerate due to the $PT$ symmetry, so that a nodal point must be at least fourfold degenerate. Under this condition and with spin splitting from strong SOC, one can expect that Weyl points should be ubiquitous at high-symmetry points. Note that the time reversal invariant momentum (TRIM) points automatically have the Kramers twofold degeneracy. However, no good candidate 2D material of this type has been identified so far.

Spin-orbit Weyl points may also appear on high-symmetry lines. A possible mechanism was reported in Ref.~\cite{Young2015Dirac}, where a pair of essential spin-orbit Weyl points are enforced by $T$ and a twofold screw rotation. The idea was demonstrated in strained monolayer GaTeI~\cite{Wu2019Hourglass}, but the material is far from ideal.

For magnetic materials, since $T$ is already broken, the constraint on $P$ can be relaxed. A large spin splitting can be achieved in a ferromagnetic materials with strong exchange interaction.
Perhaps the best example of spin-orbit Weyl points so far is in the magnetic PtCl$_3$ monolayer~\cite{You2019Two}. There, the Weyl points are located at the Fermi level without other extraneous bands and are fully spin-polarized.
We will discuss more on this material in Sec.~\ref{magtsm}.

{\color{blue}\emph{Dirac point.}} Dirac points are fourfold degenerate nodal points. By definition, it can be regarded as consisting of two superposing Weyl points with opposite chirality. In other words, the effective model at a Dirac point should be $4\times 4$ and can be expressed as
\begin{equation}\label{dirac}
  H_D=H_{W+}\oplus H_{W-},
\end{equation}
where $H_{W\pm}$ are $2\times 2$ Weyl models with opposite chirality. Clearly, the nodal point in graphene does not fulfill this definition.
When including spin, the original nodal point would be gapped by SOC, although the gap is very small ($\sim 1$ $\mu$eV) due to the weak SOC in graphene~\cite{Min2006,Yao2007}.

\begin{figure}[h]
	\centering
	\includegraphics[angle=0, width=0.45\textwidth]{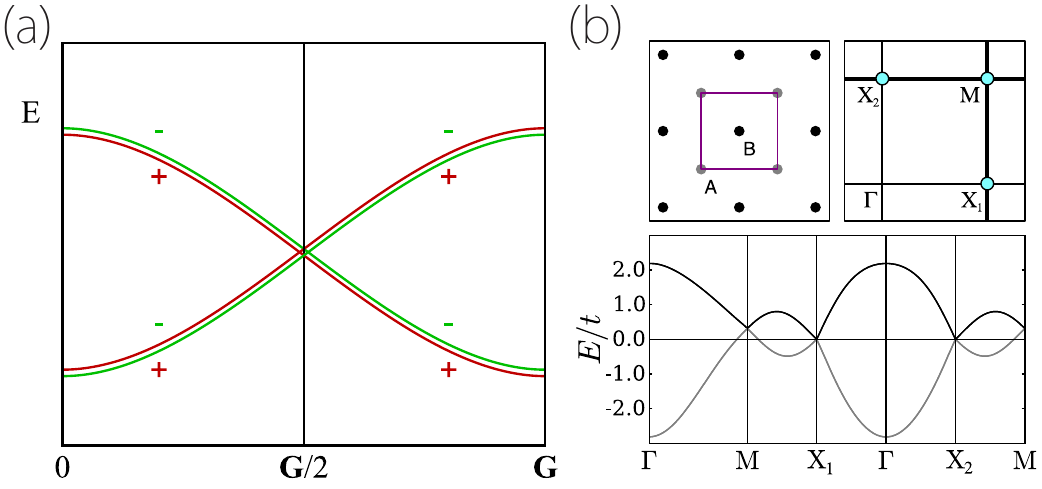}
	\caption{ (a) With $P$, $T$, and a twofold nonsymmorphic symmetry, a Dirac point can be enforced at the BZ boundary. (b) A tight-binding model which realizes three Dirac points in the BZ. Here, each band has an inherent twofold degeneracy due to $PT$. Figures adapted with permission from Ref.~\cite{Young2015Dirac}.}
	\label{Kane}
\end{figure}

Young and Kane~\cite{Young2015Dirac} first proposed a mechanism for realizing spin-orbit Dirac points in 2D. They showed that $P$, $T$, and a nonsymmorphic symmetry such as twofold screw axis or glide mirror can enforce essential Dirac points at high-symmetry points of the BZ boundary. It was shown that for 2D TSMs of this type, there must be two or three Dirac points in the BZ; single Dirac point is impossible~\cite{Young2015Dirac}.

\begin{figure}[h]
	\centering
	\includegraphics[angle=0, width=0.46\textwidth]{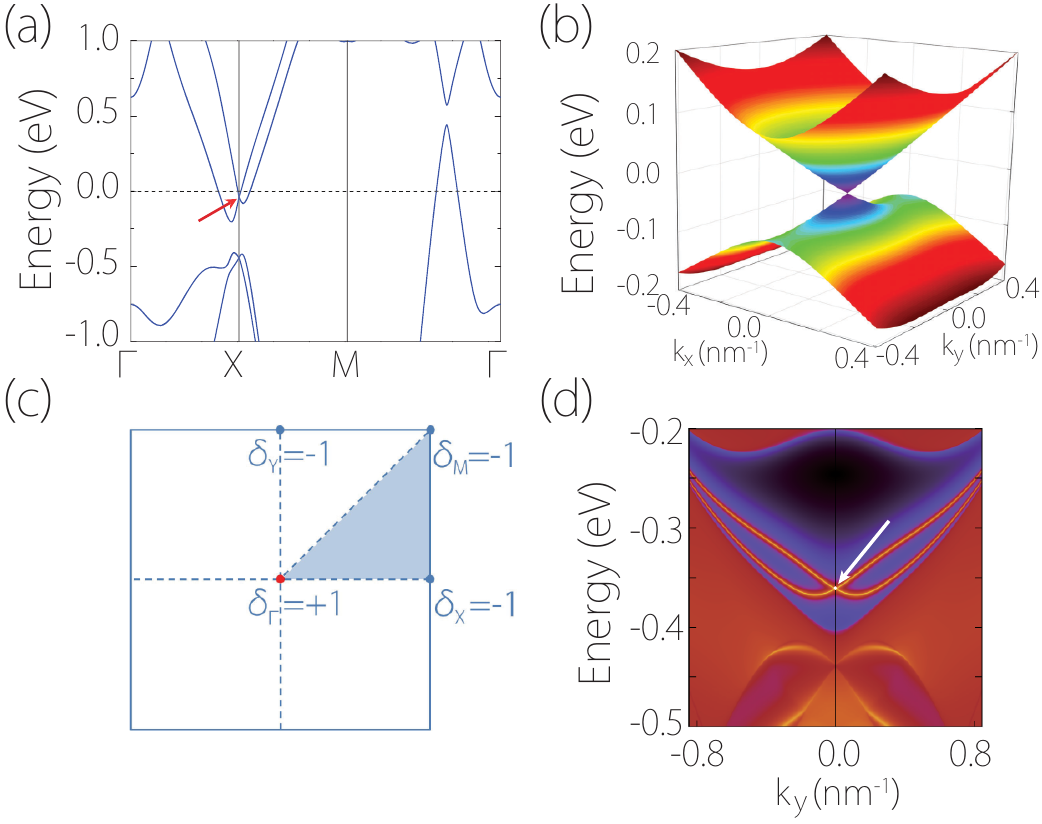}
	\caption{ 2D Dirac points in monolayer HfGeTe. (a) Band structure of ML-HfGeTe with SOC included. The red arrow indicates the spin-orbit Dirac point. (b) Energy dispersions around the Dirac point at $X$, showing anisotropic Dirac-cone spectrum. (c) Product of band parity eigenvalues at the four TRIM points, indicating a nontrivial $\mathbb{Z}_2$ invariant. Hence, monolayer HfGeTe is also a 2D $\mathbb{Z}_2$ topological metal.  (d) The edge spectrum showing a pair of spin-helical edge bands, similar to those for 2D topological insulators. They are due to the bulk $\mathbb{Z}_2$ invariant. Figures adapted with permission from Ref.~\cite{Guan2017Two}.}
	\label{hgt}
\end{figure}

The first real 2D material hosting Dirac points was predicted by Guan \emph{et al.}~\cite{Guan2017Two}. It was found that in monolayer HfGeTe-family materials, the nonsymmorphic symmetry group enforces two Dirac points at $X$ and $Y$ points, as shown in Fig.~\ref{hgt}. An example Dirac model, obtained for the Dirac point at $X$ in monolayer HfGeTe, is given by~\cite{Guan2017Two}
\begin{equation}\label{hamhgt}
  H=v_x k_x(\cos\theta\ \sigma_z\otimes \tau_z+\sin\theta\ \sigma_x\otimes\tau_z)+v_y k_y \sigma_y\otimes\tau_z,
\end{equation}
where $\tau$'s are also Pauli matrices and $\theta$ is a model parameter. It was noted that monolayer HfGeTe has spin-helical edge states [see Fig.~\ref{hgt}(d)]. They are
due to the fact that the metallic system can be continuously evolved into a 2D quantum spin Hall insulator by adjusting the local gaps. The system belongs to the $\mathbb{Z}_2$ topological metal~\cite{Pan2014}.

\begin{figure*}[tbp]
	\centering
	\includegraphics[angle=0, width=0.8\textwidth]{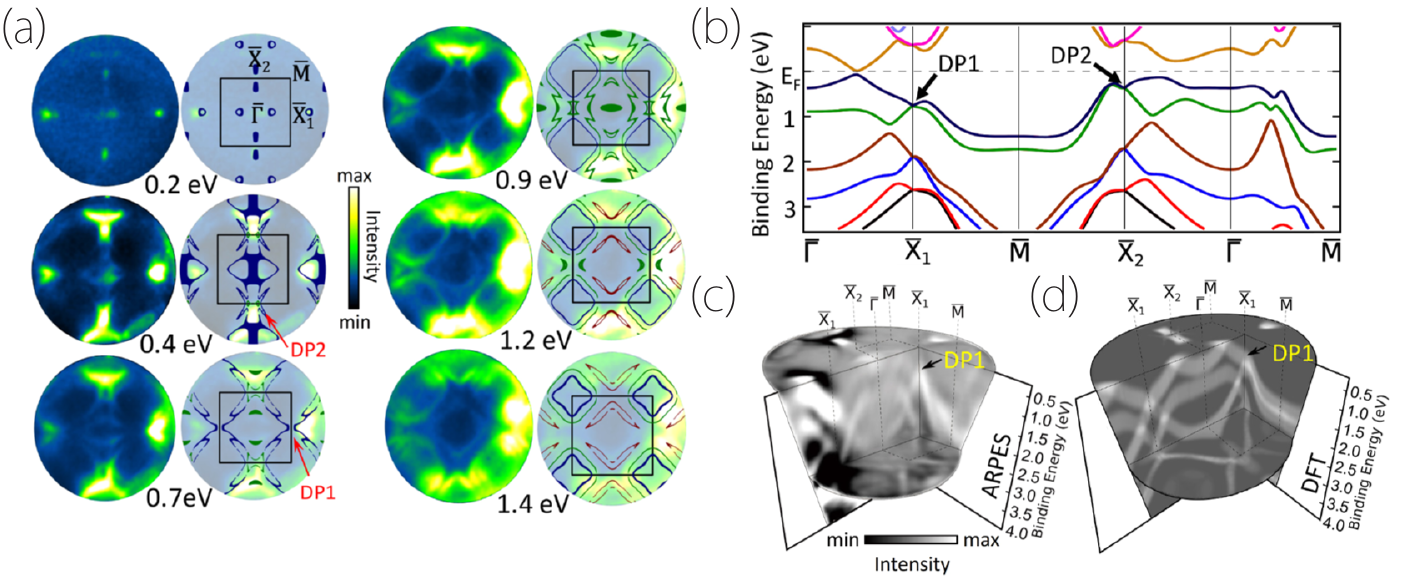}
	\caption{ Experimental detection of Dirac points in 2D $\alpha$-Bismuthene. (a) $\mu$-ARPES iso-energy contours taken at different binding energies. The locations of the Dirac points are marked by red arrows. (b) Calculated band structure. (c) 3D band representation of the ARPES result, compared to (d) the calculated 3D band contour. Figures adapted with permission from Ref.~\cite{Kowalczyk2020}. }
	\label{abi}
\end{figure*}

The first experimental detection of Dirac points in 2D was performed on $\alpha$-Bismuthene~\cite{Kowalczyk2020}. By using $\mu$-ARPES techniques, the location and the linear dispersion of the Dirac points were directly imaged, which conformed with first-principles calculation result.

It should be noted that both monolayer HfGeTe and $\alpha$-Bismuthene are not good-quality 2D Dirac TSMs. In $\alpha$-Bismuthene, the Dirac point was away from the Fermi level. In monolayer HfGeTe, the low-energy band structure is not clean; it is a metal rather than a semimetal. Hence, it remains a challenge to search for a suitable 2D Dirac TSM.

\begin{figure}[h]
	\centering
	\includegraphics[angle=0, width=0.48\textwidth]{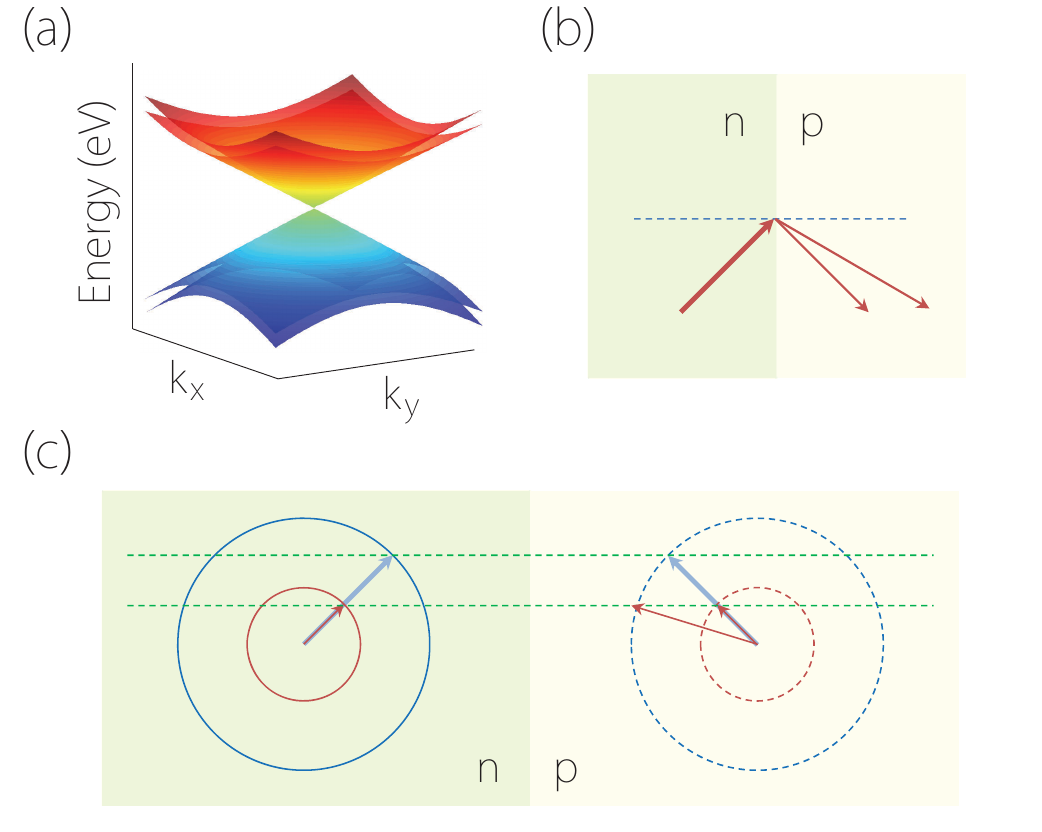}
	\caption{ (a) Energy dispersion around a birefringent Dirac point. (b) Double refraction occurs for an incident particle beam. (c) shows a n-p junction setup in TSMs with birefringent Dirac point.  Figures adapted with permission from Ref.~\cite{Chen2017}.}
	\label{cab}
\end{figure}

The original concept of a Dirac point requires each of the two crossing bands has twofold degeneracy, as consistent with the concept of Dirac fermions. In materials, this requirement is usually fulfilled by the $PT$ symmetry. If we relax this requirement and define any fourfold points as Dirac, we could have so-called birefringent Dirac points~\cite{Kennett2011,Roy2012,Komeilizadeh2014}. The naming is from the analogy with optics: Here, the twofold degeneracy of each band is lifted, so around each Dirac point there are two equi-energy surfaces, which give rise to two refracted beams for a given incident particle beam (see Fig.~\ref{cab}). In 3D materials, the birefringent Dirac points have been discussed, e.g., in CaAgBi family materials~\cite{Chen2017} and 3D honeycomb carbon~\cite{Hu2019}. In 2D, Jin \emph{et al.}~\cite{Jin2020} proposed that such point can be enforced by $T$ and two glide mirrors, and predicted its realization in monolayer SbSSn. Notably, due to the $P$ breaking, there can be nonzero Abelian Berry curvature around such a point, which is impossible for $PT$-invariant systems.

Similar birefringent Dirac points have also been reported in spinless 2D systems. Ref.~\cite{Damljanovic2017} listed three layer groups that host such points, and a hypothetical phosphorous 2D structure was proposed, which hosts the point near the Fermi level.

Dirac points can also exist in 2D magnetic materials. This will be discussed later in Sec.~\ref{magtsm}.

\begin{figure}[h]
	\centering
	\includegraphics[angle=0, width=0.48\textwidth]{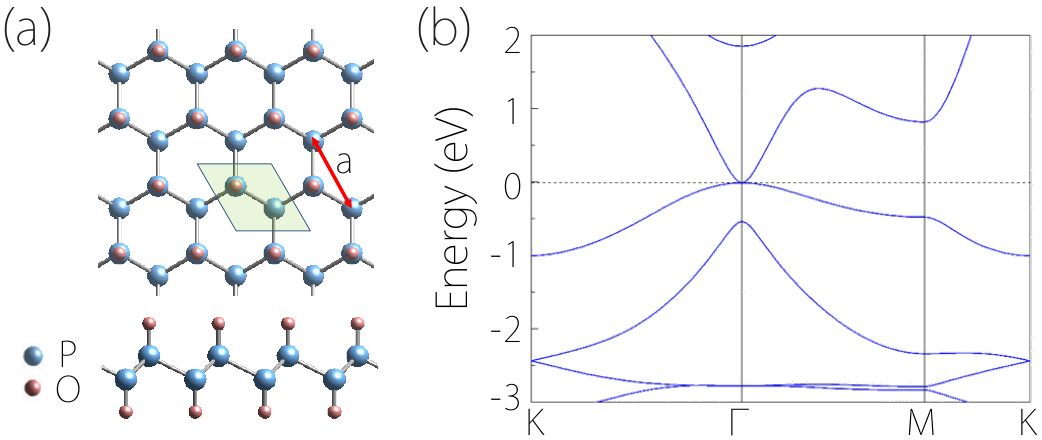}
	\caption{(a) Top and side views of monolayer blue phosphorene oxide.  (b) Electronic band structure for blue phosphorene oxide under 4\% strain, showing a double Weyl point at Fermi level. Figures adapted with permission from Ref.~\cite{Zhu2016Blue}. }
	\label{bpo}
\end{figure}

{\color{blue}\emph{Double Weyl point.}} Double Weyl points (or quadratic Weyl points) are Weyl points with a quadratic band energy splitting, namely,
\begin{equation}
  \Delta E(\bm k)=E_1(\bm k)-E_2(\bm k)\sim k^2,
\end{equation}
where $\Delta E$ is the energy splitting between the two bands. Note that in $AB$-stacked bilayer graphene, the low-energy dispersion is approximately quadratic at $K$ and $K'$ points when taking into account only the nearest-neighbor interlayer hopping~\cite{Neto2009}. Nevertheless, the quadratic point is unstable when including more hopping terms, which split it into four linear Weyl points (so-called trigonal warping effect).

The first stable double Weyl point was reported in strained blue phosphorene oxide, which is a spinless system~\cite{Zhu2016Blue}. As shown in Fig.~\ref{bpo}, the point occurs at $\Gamma$ and is protected by the $D_{3d}$ symmetry. The generic model for such a point is given by
\begin{equation}
  H=A k^2 +B \begin{bmatrix}
 0 & {k}_{-}^2 \\
 k_{+}^2 & 0
\end{bmatrix},
\end{equation}
where $k_\pm=k_x\pm ik_y$, $A$ and $B$ are real model parameters. The point has a Berry phase of $2 \pi$, which is trivial when modulo $2\pi$. Nevertheless, the double Weyl fermions do exhibit distinct properties. For example, a universal optical absorbance was predicted in Ref.~\cite{Zhu2016Blue} that for low-frequency optical excitations ($<0.5$ eV for blue phosphorene oxide), the absorbance $A(\omega)=\pi \alpha\simeq 2.3\%$  is a universal value.
Here, $\alpha=e^2/(\hbar c)\simeq 1/137$ is the fine-structure constant.

\begin{figure}[h]
	\centering
	\includegraphics[angle=0, width=0.485\textwidth]{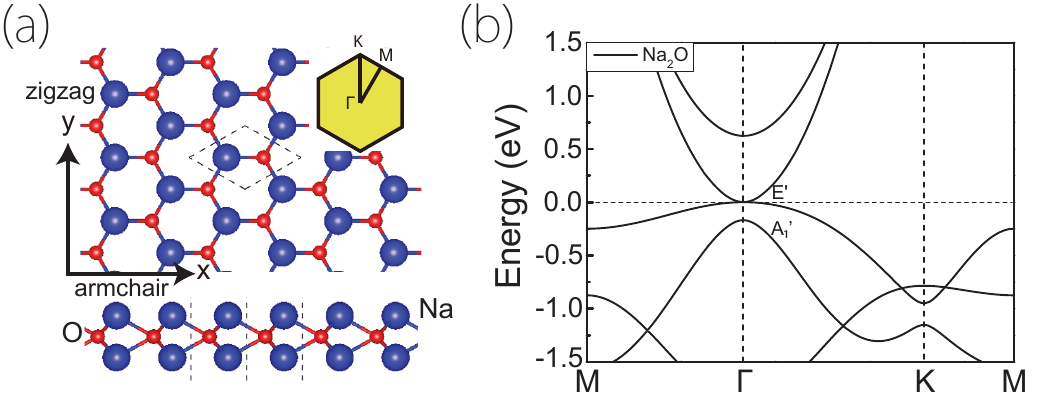}
	\caption{ (a) Top and side view of monolayer Na$_2$O. (b) Band structure of monolayer Na$_2$O at equilibrium state, showing a double Weyl point at Fermi level. Figures adapted with permission from Ref.~\cite{Hua2020tunable}. }
	\label{na2o}
\end{figure}

Later, double Weyl points were also found in monolayer Mg$_2$C~\cite{Wang2018} and monolayer Na$_2$O and K$_2$O~\cite{Hua2020tunable}. For Mg$_2$C and K$_2$O, the appearance of double Weyl points require lattice strain, whereas for Na$_2$O, the point exists in the equilibrium state. Therefore, monolayer Na$_2$O seems to be a good platform to explore the double Weyl fermions.

The double Weyl points reported so far are all spinless. It is currently not clear whether such points can exist in systems with SOC. This is an interesting question to be explored.

{\color{blue}\emph{Other cases.}} Here, we mentioned two other types of nodal points in 2D discussed in literature: the semi-Dirac point and the pseudospin-1 point. It should be noted that they are \emph{not stable}, i.e., they are not symmetry protected and their existence requires fine tuning of system parameters.

The semi-Dirac point, which according to our naming convention should be called a semi-Weyl point, was first reported in certain VO$_2$/TiO$_2$ nano-structures~\cite{Pardo2009}. It has linear band splitting along one direction and quadratic splitting along the other direction. A generic effective model is given by
\begin{equation}
  H=vk_x\sigma_z+\frac{k_y^2}{2m}\sigma_x.
\end{equation}
As mentioned, such points are not protected. It in fact represents a critical state at the transition when two unpinned linear Weyl points are merged and about to open a gap.

\begin{figure}[h]
	\centering
	\includegraphics[angle=0, width=0.48\textwidth]{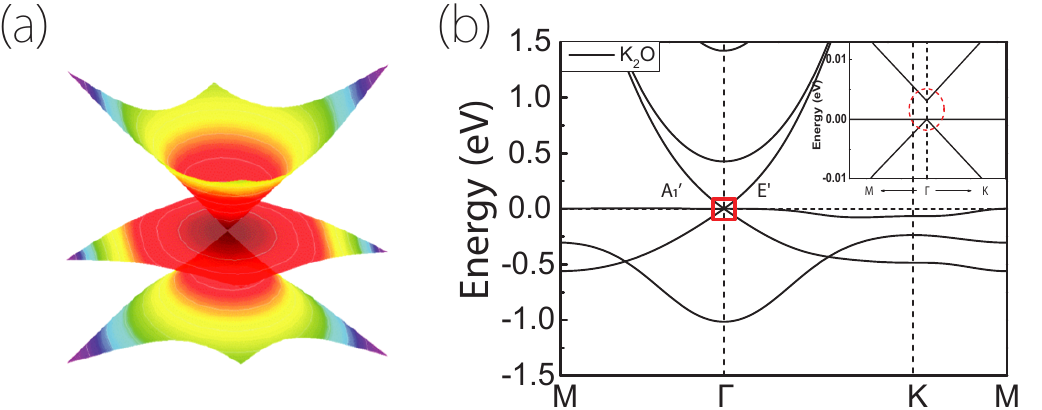}
	\caption{ (a) Energy dispersion around a pseudospin-1 point. (b) Electronic band structure for K$_2$O at equilibrium state. Around the Fermi level, the bands are very close to forming a pseudospin-1 point.  Figures adapted with permission from Ref.~\cite{Zhu2016Blue,Hua2020tunable}.}
	\label{sp1}
\end{figure}

The pseudospin-1 point is a triply degenerate point. Its spectrum consists of a Weyl cone and a flat band intersecting the Weyl point, as shown in Fig.~\ref{sp1}. It was first discussed in cold-atom systems and some molecular patterned crystal surface~\cite{Shen2010,Goldman2011,Paavilainen2016}. The first 2D material example showing this feature is the strained blue phosphorene oxide (which is spinless)~\cite{Zhu2016Blue}. It represents a critical state at a semiconductor-semimetal transition. Its effective model is given by
\begin{equation}
  H=v\bm k\cdot \bm S,
\end{equation}
where $S$'s are two $3\times 3$ purely imaginary Gell-Mann matrices. Such points attracted interest, because the pseudospin-1 fermions have remarkable effects, such as super-Klein tunneling~\cite{Shen2010,Urban2011}, supercollimation~\cite{Fang2016}, and super Andreev reflection~\cite{Feng2020}. We will not elaborate on these effects here. The interested readers may consult the listed references. The pseudospin-1 point was later also reported in strained monolayer Mg$_2$C~\cite{Wang2018}, monolayer Na$_2$O and K$_2$O~\cite{Hua2020tunable}. It is noted that monolayer K$_2$O in its equilibrium state is very close to the critical state [see Fig.~\ref{sp1}(b)], suggesting it as a good platform to study pseudospin-1 fermions.

\section{\label{tnlsm} 2D nodal-line TSM}

{\color{blue}\emph{Spinless Weyl line.}} Weyl lines are common for 2D materials which preserve a horizontal mirror plane. This is because in such cases, each band have a definite parity under mirror, and the crossing between two bands with different parity will generically form a Weyl line in the 2D BZ.
These lines have been found in many material examples, such as Hg$_3$As$_2$~\cite{Lu2017Two}, Ca$_2$As~\cite{Niu2017Two}, PdS~\cite{Jin2017The}, C$_9$N$_4$~\cite{Chen2018Prediction}, p-IVX$_2$~\cite{Zhang2018From}, and honeycomb borophene oxide~\cite{Zhong2019}.

\begin{figure}[h]
	\centering
	\includegraphics[angle=0, width=0.48\textwidth]{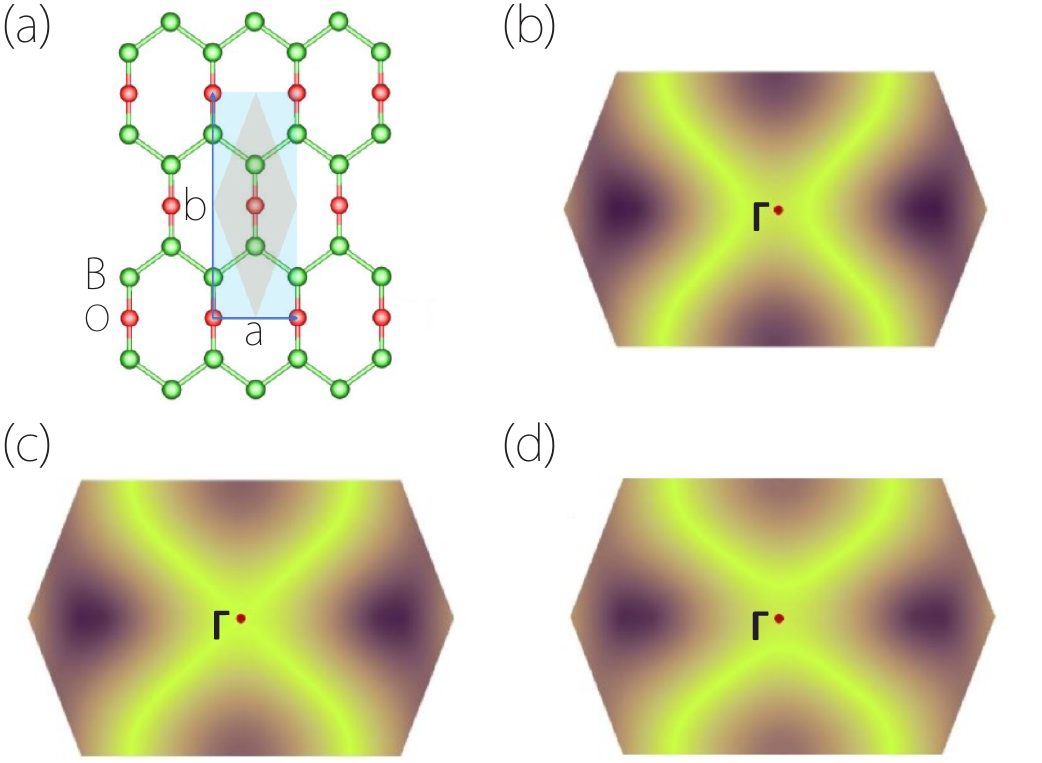}
	\caption{ Nodal loop transformation in honeycomb borophene oxide. (a) Lattice structure of honeycomb borophene oxide. Nodal-loop profile under (b) 5\% compressive strain, (c) 1.45\% compressive strain, and (d) 5\% tensile strain along the $a$ direction. In the process, a single ring splits into two loops traversing the BZ. Figures adapted with permission from Ref.~\cite{Zhong2019}. }
	\label{loops}
\end{figure}

This type of lines are in a sense unpinned: they are not restricted to high-symmetry paths and their shapes may be deformed under perturbation. As mentioned in Sec.~\ref{basicConcept}, we may characterize the topology of a loop in 2D BZ by two integer numbers, corresponding to the number of times the loop traverses the BZ in the two directions~\cite{Li2017}. Mathematically, this characterization derives from the fundamental homotopy group of the two torus $\pi_1(\mathbb{T}^2)=\mathbb{Z}\times\mathbb{Z}$. In this sense, a nodal ring around a high-symmetry point [Fig.~\ref{patt}(a)] and a nodal loop traversing the BZ [Fig.~\ref{patt}(b)] are distinct. Physically, this distinction manifests in the fact that under symmetry-preserving perturbations, the former can continuously shrink to a point, whereas the latter cannot (by itself); to make the shrinkage, the latter one must do it with a partner. An example is shown in the honeycomb borophene oxide~\cite{Zhong2019}. As shown in Fig.~\ref{loops}(b-d), under strain which preserves the horizontal mirror, the two loops in Fig.~\ref{loops}(b) can merge into a single ring (which may then shrink to a point and annihilate).

\begin{figure}[h]
	\centering
	\includegraphics[angle=0, width=0.48\textwidth]{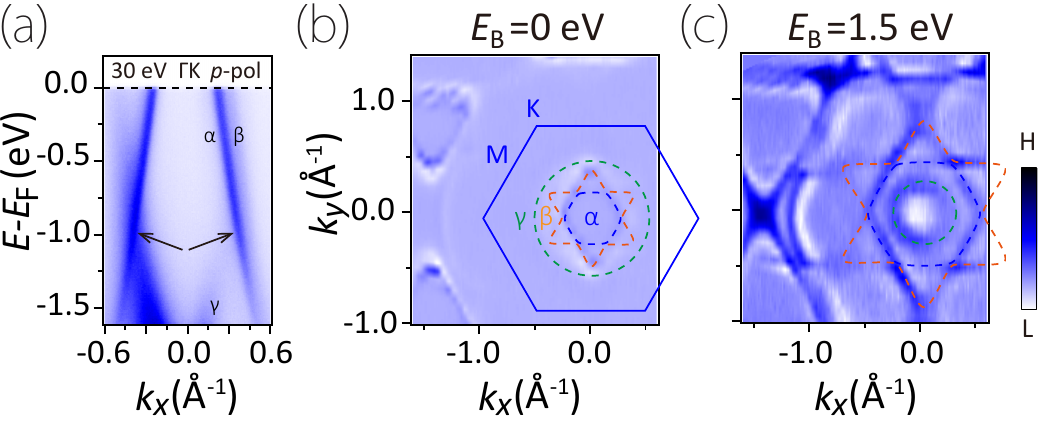}
	\caption{ ARPES experiment on monolayer Cu$_2$Si, which probes the mirror-protected spinless Weyl lines in the band structure. Figures adapted with permission from Ref.~\cite{Feng2017}. }
	\label{cu2si}
\end{figure}

Experimentally, nodal lines in monolayer Cu$_2$Si~\cite{Feng2017} and CuSe~\cite{Gao2018Epitaxial} have been detected by ARPES measurement. The nodal lines in these two examples belong the mirror protected spinless Weyl lines.

\begin{figure}[h]
	\centering
	\includegraphics[angle=0, width=0.48\textwidth]{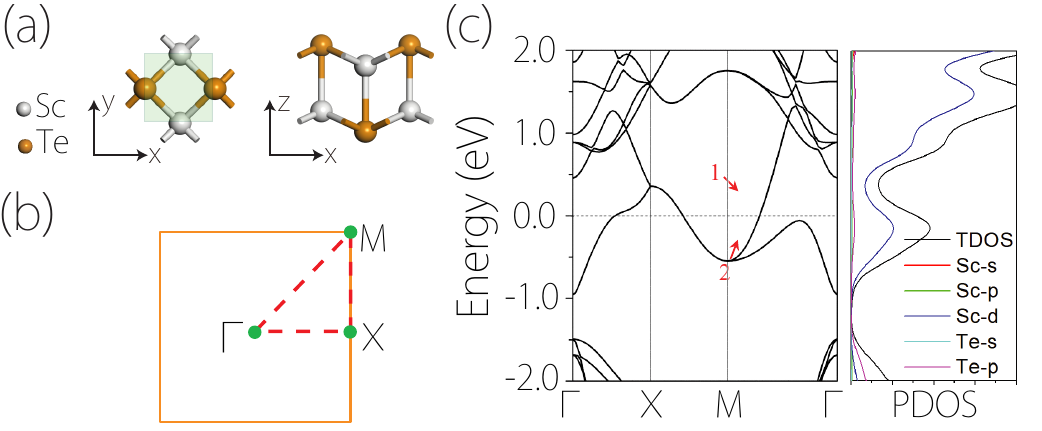}
	\caption{ (a) Lattice structure of monolayer ScTe. (b) Essential Weyl lines exist at the entire BZ boundary (yellow). (c) Calculated band structure and PDOS of monolayer ScTe in the absence of SOC. Figures adapted with permission from Ref.~\cite{Guo2020a}.}
	\label{ScTe}
\end{figure}

Another type of spinless Weyl lines are enforced by nonsymmorphic symmetries at high-symmetry paths of the BZ. Consider a twofold screw rotation along $x$. It can be shown that the combined operation $(TS_x)$ satisfies
\begin{equation}\label{TS}
  (TS_x)^2=-1
\end{equation}
along the $k_x=\pi$ path at the BZ boundary. It follows that each band on this path must have a Kramers like twofold degeneracy, corresponding to a Weyl line on this path. This type of lines have been reported in monolayer ScTe-family materials~\cite{Guo2020a}. Notably, when there are two in-plane screw axes $S_x$ and $S_y$, nodal line will form at the entire BZ boundary. This is the case in monolayer ScTe, and can be viewed as a kind of nodal cage. It was shown in Ref.~\cite{Yu2019Circumventing} that its counterpart in 3D can coexist with a single Weyl point in BZ, offering a route to circumvent the no-go theorem.

{\color{blue}\emph{Spin-orbit Weyl line.}} The two types of Weyl lines discussed above also find counterparts in spin-orbit systems, because the corresponding symmetry protection works for both cases.

For the first type, the mirror eigenvalue is still well defined for each band. Hence a Weyl line protected by the horizontal mirror symmetry can still be stabilized. To have a Weyl line, the two bands that form the line should have sufficiently large spin splitting. This also requires that the $PT$ symmetry must be broken.

For the second type, one notes that the symmetry relation in Eq.~(\ref{TS}) remains valid for a spinful particle, because although $T^2=-1$ has a sign flip, $(S_x)^2$ also brings an extra minus sign due to the $2\pi$ rotation of spin-1/2. As a result, the twofold degeneracy persists on the $k_x=\pi$ path. And again, it is necessary to have broken $PT$ symmetry. Otherwise, the $k_x=\pi$ would not correspond to a Weyl line but just one doubly degenerate band. This analysis in fact parallels that for the Class-II nodal surfaces in 3D systems with SOC~\cite{Wu2018a}.

\begin{figure*}[htbp]
	\centering
	\includegraphics[angle=0, width=0.9\textwidth]{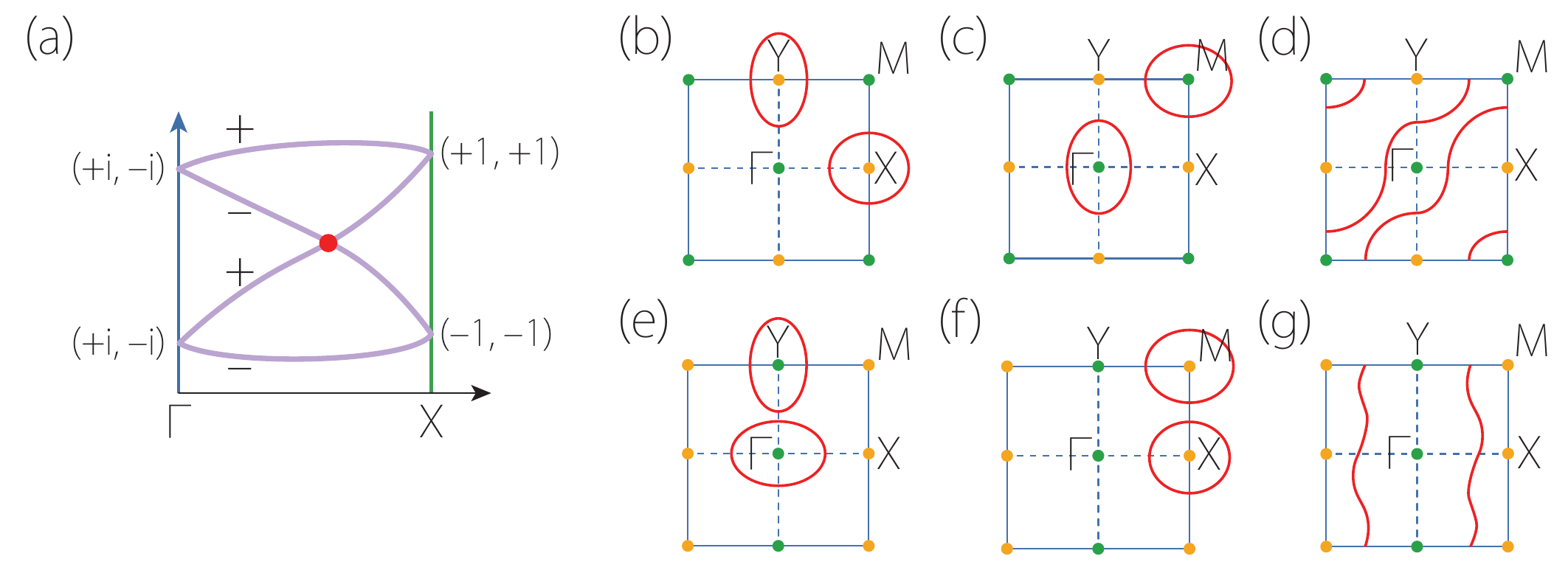}
	\caption{ (a) Schematic figure showing the hourglass dispersion along a path connecting two TRIM points. (b-d) Possible patterns of spin-orbit Weyl loops in the 2D BZ for the glide mirror $\{M_z|\frac{1}{2}\frac{1}{2}\}$. (e-g) Possible patterns of spin-orbit Weyl loops for the glide mirror $\{M_z|\frac{1}{2}0\}$.  Figures adapted with permission from Ref.~\cite{Wu2019Hourglass}.}
	\label{hourglass}
\end{figure*}

It should be mentioned that under the first type, there is a sub-type of essential spin-orbit Weyl lines. These lines are protected and further enforced by a horizontal glide mirror symmetry. And each point on the line corresponds to the neck point of an hourglass type dispersion. The existence of such lines was first noticed in Ref.~\cite{Young2015Dirac}, which is derived from the spin-orbit Dirac points after breaking the inversion symmetry. Wu \emph{et al.}~\cite{Wu2019Hourglass} systematically studied the symmetry condition and the topology of this type of lines. Fig.~\ref{hourglass} illustrate the possible line profiles.
Materials monolayer GaTeI~\cite{Wu2019Hourglass} and Bi/Cl-SiC(111)~\cite{Wang2019Hourglass} were predicted to host such lines, but in both examples, the lines are away from the Fermi level.

So far, a good candidate material with spin-orbit Weyl lines is still lacking.

{\color{blue}\emph{Dirac line.}} Fourfold degenerate Dirac lines in 2D was recently proposed in Ref.~\cite{Guo2020,Cui2020}. It was shown that with SOC, the symmetries $P$, $T$, a screw axis $S_{2x}$ and a vertical glide mirror $G_y$ can enforce a Dirac line on the $k_y=\pi$ path at the BZ boundary. Such a line was predicted in the so-called brick phase tri-atomic layer Bi(110), which was successfully grown on black phosphorous substrate in experiment~\cite{Cui2020}.

\section{\label{magtsm} 2D Magnetic TSM}

We arrange a separate section for discussing magnetic TSMs, because the presence/absence of the $T$ symmetry has important impacts on topological classifications. The spatial operations will also act on the magnetic moments which typically lowers the symmetry, and one has to deal with magnetic space groups. In addition, magnetic TSMs offer possibility to study the interplay between magnetism and topological fermions. For example, the band topology may be controlled by varying temperature across magnetic phase transitions or by tuning the magnetization direction.
From a practical point of view, the most desired case is to have a topological half semimetal, such that the topological fermions can be 100\% spin polarized, which will be useful for spintronics applications. The terminology initially appeared in Ref.~\cite{Liu2015Multiple} for the semihydrogenated germanene, where a double Weyl point at Fermi level occurs in a single spin channel and the other channel is gapped.

To have stable magnetism in 2D materials, strong magnetic anisotropy is required to fight against the enhanced fluctuations with reduced dimensionality. Typically, the magnetic anisotropy is dominated by SOC effects. Hence, in the following discussion, we will focus on the TSM states that are robust under SOC.

\begin{figure}[h]
	\centering
	\includegraphics[angle=0, width=0.48\textwidth]{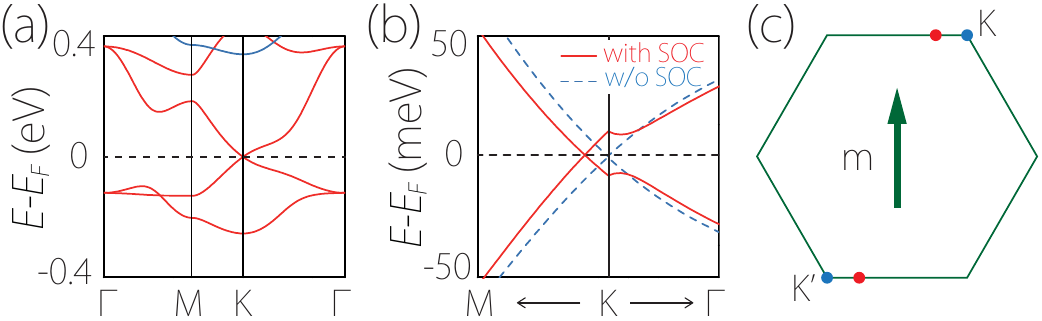}
	\caption{ (a) Band structure of ferromagnetic monolayer PtCl$_3$ without SOC. The red and blue bands are for spin-up and spin-down channels, respectively. (b) Enlarged view of the band structure around the Weyl point. The red solid (blue dashed) lines are for the bands with (without) SOC. (c) Two Weyl points are located at $K/K'$ points without SOC (blue points), and they are shifted along the $x$ direction on the mirror-invariant line after considering SOC (red points). Figures adapted with permission from Ref.~\cite{You2019Two}.}
	\label{ptcl3}
\end{figure}

{\color{blue}\emph{Weyl point.}} Similar to the nonmagnetic case, magnetic Weyl points could exist at high-symmetry points, corresponding to 2D irreducible representations of the little group; or formed by crossing between two bands of different symmetry on high-symmetry paths. A good example is monolayer PtCl$_3$~\cite{You2019Two}. The system is ferromagnetic with in-plane magnetization direction. The magnetism preserves a vertical mirror, and a pair of Weyl points are protected by this mirror symmetry on the mirror-invariant path, robust under SOC, as shown in Fig.~\ref{ptcl3}. The advantages of the material are clear. Due to electron filling, the Weyl points sit exactly at the Fermi energy, so that the Fermi surface is clean and consists of only the two Weyl points. In addition, the low-energy bands belong to a single spin channel, while the other channel has a gap $\sim 1$ eV. Hence, the material is a good candidate of a Weyl half semimetal.

{\color{blue}\emph{Dirac point.}} So far, Dirac points have been found in antiferromagnetic 2D systems with $PT$ symmetry. Here, $PT$ helps to enforce a double degeneracy for each band, and it requires the magnetic ordering to be antiferromagnetic. The possibility was first pointed out by Wang~\cite{Wang2017a} and by Young and Wieder~\cite{Young2017Filling}. Ref~\cite{Young2017Filling} proposed one sufficient condition for achieving magnetic Dirac points, which involves two spatial symmetries and a special ``nonsymmorphic" time reversal symmetry $\tilde{T}=\{T|\bm t\}$, with $\bm t$ a fractional lattice translation. The resulting Dirac point is at a high-symmetry point and it is possible to have a single Dirac point in BZ. Meanwhile, Ref~\cite{Wang2017a} gave a general analysis for Dirac points in 2D systems with $PT$ symmetry. It further identified possible pairs of Dirac points on high-symmetry paths, protected by nonsymmorphic symmetries.

A material example, the FeSe monolayer, was proposed in Ref.~\cite{Young2017Filling}. However, its Dirac point is below the Fermi level by more than 0.4 eV, and the energy splitting between the two crossing bands is very small along one direction. Another example is monolayer TaCoTe$_2$~\cite{Li2019Two}. Both type-I and type-II Dirac points were found in the antiferromagnetic state, depending on the N\'{e}el vector direction. Nevertheless, these points are located around $-0.25 $ eV in the valence band. Currently, we do not have a good magnetic Dirac semimetal material.

{\color{blue}\emph{Weyl line.}} The two protection mechanisms for spin-orbit Weyl lines discussed in Sec.~\ref{tnlsm} can be extended to magnetic systems. The first type, namely, the protection by a horizontal mirror, is the mostly encountered case. This horizontal mirror is compatible with ferromagnets and antiferromagnets with out-of-plane magnetic moments. A glide horizontal mirror may also be preserved in certain antiferromagnets with in-plane moments. A Weyl line emerges when two bands with opposite mirror eigenvalues cross each other.

The second type with $TS_{x}$ symmetry (and with broken $PT$) also works, since $TS_{x}$ itself is a kind of magnetic symmetry and is compatible with certain antiferromagnetic configuration.

\begin{figure}[h]
	\centering
	\includegraphics[angle=0, width=0.48\textwidth]{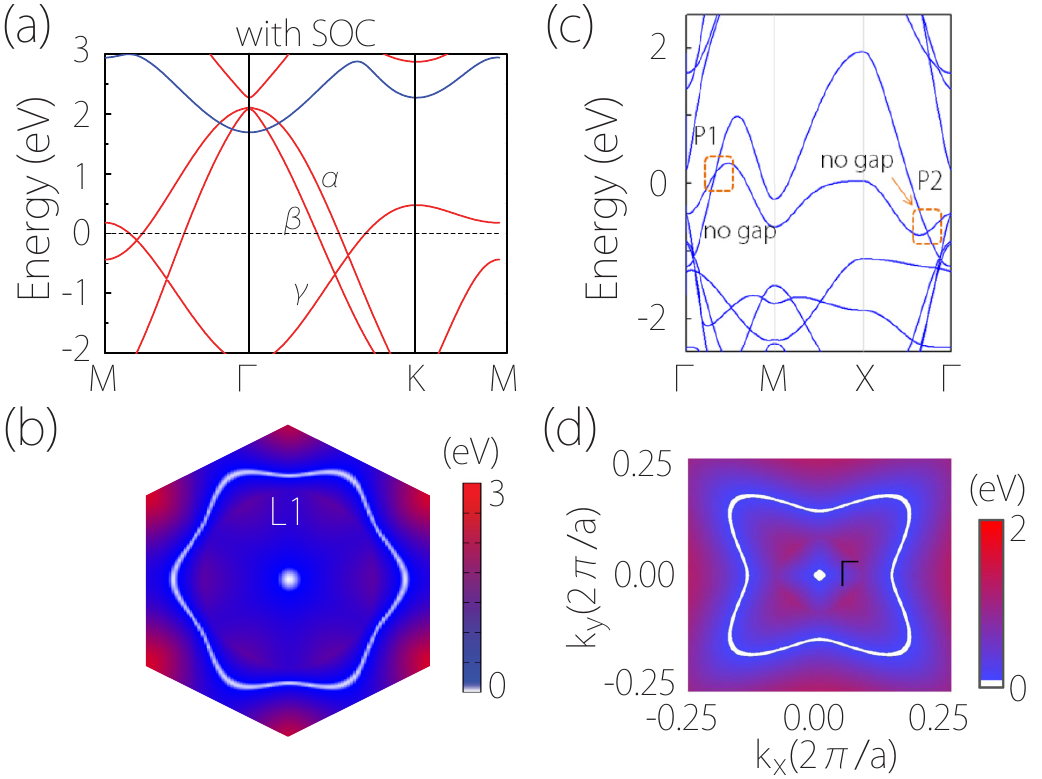}
	\caption{ (a) Band structure of ferromagnetic monolayer MnN with SOC. The crossing of $\alpha,~\beta$ and $\gamma$ bands forms two nodal loops near Fermi level. (b) shows the outer loop in monolayer MnN. (c) Band structure of monolayer CrN with SOC. (d) shows the hybrid nodal loop in monolayer CrN. Figures adapted with permission from Ref.~\cite{Wang2019Two,He2020}. }
	\label{MnN}
\end{figure}

So far, magnetic Weyl-line material candidates have been proposed for the first type, e.g., in monolayer MnN~\cite{Wang2019Two}, CrN~\cite{He2020}, AgN~\cite{Li2020}, CsS~\cite{Zhou2019Fully}, CoSe~\cite{Tai2020}, InC~\cite{Jeon2018}, and K$_2$N~\cite{Jin2020b}. Fig.~\ref{MnN} shows the magnetic Weyl lines in monolayer MnN and CrN. Here, the lines are close to the Fermi level, and the low-energy bands are fairly clean. No material candidate of the second type has been reported yet. Experimentally, the Weyl lines in monolayer GdAg$_2$ (which is ferromagnetic) has been probed by ARPES~\cite{Feng2019Discovery}.

\section{\label{tes} Topological edge states}

Topological boundary states come with protection from topological invariants defined in the bulk. For example, in 3D Weyl semimetals, the surface Fermi arcs are protected by bulk Chern numbers~\cite{Wan2011Topological}; and the drumhead surface states in nodal line semimetals are from the quantized Berry phase $\nu$~\cite{Yang2014b,Weng2015}. For 2D systems, possible bulk invariants are 2D invariants defined for the whole BZ (such as Chern number and $\mathbb{Z}_2$ invariant for quantum spin Hall state) and 1D invariants defined on a closed loop (such as Berry phase). For 2D TSMs, ideally, the spectrum is gapless, and we do not have a well defined 2D invariant. One may note that below Eq.~(\ref{hamhgt}), we mentioned the spin-helical edge states due to 2D bulk $\mathbb{Z}_2$ invariant in monolayer HfGeTe~\cite{Guan2017Two}. However, in such case, the edge states have no direct connection to the Dirac points. In other words, they do not represent a feature (or a consequence) of the bulk nodal structure.

\begin{figure}[h]
	\centering
	\includegraphics[angle=0, width=0.5\textwidth]{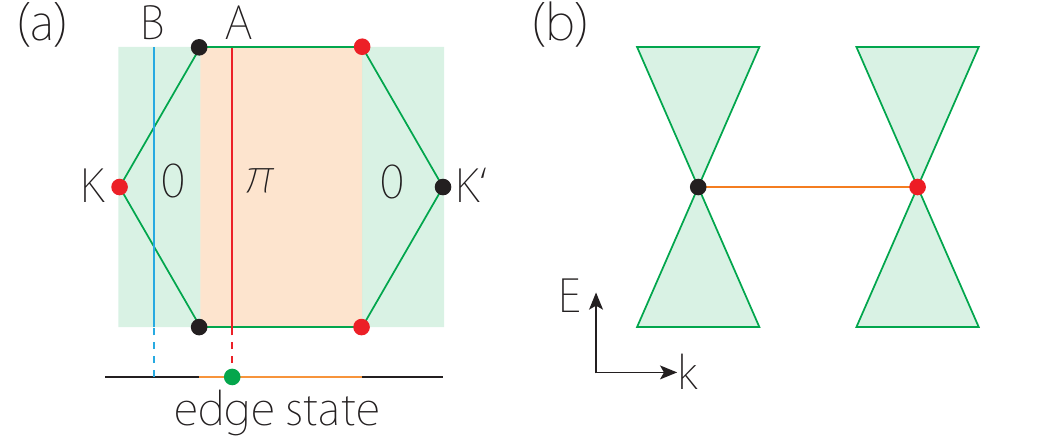}
	\caption{Zak phase and protected edge states. (a) For graphene, the Zak phase is nontrivial in the red shaded region between the two Weyl points and is trivial outside.  (b) This leads to the flat edge band for the zigzag edge of graphene. }
	\label{zak}
\end{figure}

A connection can be established between edge flat band and bulk nodal points with nontrivial 1D invariant. A good example is graphene. The two Weyl points in graphene carry a $\pi$ Berry phase $\nu=1$ (or $\nu_C=\pm 1$ if counting the chiral symmetry). This guarantees a $\pi$ phase difference between the two paths shown in Fig.~\ref{zak}. Here, path A has a nontrivial $\nu=1$, so on the zigzag edge, there exists a flat edge band connecting the projections of the two Weyl points. It should be noted that although called an edge flat band, in practice the band is generally dispersive, and the dispersion can be controlled by modifying the boundary~\cite{Yao2009}. Boundary modification may also change the BZ region for the appearance of the edge band. 

With the above analysis, we see that the chances to have topological boundary states in 2D TSMs is much reduced compared to 3D. Particularly, for 2D nodal-line TSMs, they typically do not possess protected edge states, or even if they do, the boundary states would have no connection with the nodal lines.

\section{\label{outlook} Outlook}

The field of 2D TSMs is under rapid development. The important problems to be address in the subsequent research include the following.

First, we have introduced a variety of nodal structures where the classification of 2D TSMs is based upon. Are they complete? Is there any new class of 2D TSMs? This is a fundamental open question that needs to be addressed.

Second, regarding materials, for many classes, we do not yet have a good material candidate.
The standards for a good TSM material have been given in Sec.~\ref{basicConcept} (which would disqualify many proposed examples in literature). And graphene is a paradigmatic example of a good TSM that one should always make reference to. The lack of good materials severely hinders the research in this field. Many interesting physical effects were predicted by assuming the ideal case, which is not met in the candidate materials. For example, in studies on 3D Weyl semimetal materials, the low-energy bands for currently available materials are far from ideal: the linear window is very small, there are extraneous bands nearby, and the existence of many Weyl points forming clusters and leading to strong scattering within a cluster. This poses difficulty in isolating effects that are truly from the Weyl points and in interpreting experimental results.

We feel that considering material realizations, 2D TSMs may have advantages over 3D. For 3D, extensive computational studies have been performed on the database of existing 3D materials~\cite{Zhang2019Catalogue,Vergniory2019A,Tang2019Comprehensive}. Despite a large number of TSMs identified, good ones are quite rare (perhaps the best are still the 3D Dirac semimetals Na$_3$Bi~\cite{Wang2012} and Cd$_3$As$_2$~\cite{Wang2013}), and it is fair to say that there is no one that has quality comparable to graphene. In comparison, for 2D, we have the hope to design materials with no 3D counterparts. In addition, 2D materials are much more tunable than 3D ones. Non-ideal TSMs may be salvaged by proper material engineering.

Finally, we still need to understand the physical properties of the various 2D TSMs. We need to understand what are their distinct features and what could be useful for applications. Especially, we do not know what are the special effects of 2D nodal-line semimetals. And research needs to go beyond simple models and work more closely with realistic materials.

\begin{acknowledgments}
The authors thank D. L. Deng for valuable discussions. This work is supported by the Singapore Ministry of Education AcRF Tier 2 (Grant No. MOE2019-T2-1-101).
\end{acknowledgments}

%%%%%%%%%%%%%%%%%%%%%%%%%%%%%
%   Appendix end
%%%%%%%%%%%%%%%%%%%%%%%%%%%%%

% \bibliographystyle{cpb}
\bibliographystyle{apsrev4-1}
\bibliography{TSM2D}
	
\end{document}